\documentstyle[aps,prl,multicol,epsf]{revtex}
\def\be{\begin{equation}}
\def\ee{\end{equation}}
\newcommand{\exval}[1]{\mbox{$\langle \, {#1}\, \rangle$}}
\def\bea{\begin{eqnarray}}
\def\eea{\end{eqnarray}}

\def\ra{\rangle}
\def\la{\langle}
\def\two{\hbox{$\bigcirc$\hspace{-0.28cm}{\em 2}}\hspace{0.25cm}}

\begin{document}
\title{Transport in the $XX$ chain at zero temperature:\\
Emergence of flat magnetization profiles}

\author{T. Antal, Z. R\'acz, A. R\'akos}

\address{Institute for Theoretical Physics,
E\"otv\"os University\\
1117 Budapest, P\'azm\'any s\'et\'any 1/a, Hungary}

\author{G. M. Sch\"utz}

\address{Institut f\"ur Festk\"orperforschung,
Forschungszentrum J\"ulich \\
52425 J\"ulich, Germany}
\date{\today}
\vskip 1truecm

\maketitle

\begin{abstract}
{We study the connection between magnetization transport and
magnetization profiles in zero-temperature $XX$ chains.
The time evolution of the transverse magnetization, $m(x,t)$,
is calculated using an inhomogeneous initial state that is the ground state
at fixed magnetization but with $m$ reversed from
$-m_0$ for $x<0$ to $m_0$ for $x>0$.
In the long-time limit, the magnetization
evolves into a scaling form $m(x,t)=\Phi(x/t)$ and the profile develops
a flat part ($m=\Phi=0$) in the $|x/t|\le c(m_0)$ region.
The flat region shrinks to zero if $m_0\to 1/2$ while
it expands with the maximum velocity, $c_0=1$, for
$m_0\to 0$. The states emerging in the scaling limit are compared
to those of a homogeneous system where the same magnetization
current is driven by a bulk field, and we
find that the expectation values of various quantities
(energy, occupation number in the fermionic representation) agree
in the two systems.}~
\end{abstract}

\begin{multicols}{2}
\narrowtext

%%%%%%%%%%%%%%%%%%%%%%%%%%%%%%%%%%%%%%%%%%%%%%%%%%%%%%%%%%%%%%%%%%%%%
\section{Introduction}

Transport in integrable systems is anomalous.
This means that a harmonic lattice (or a
transverse Ising chain) cannot support an
internal thermal gradient when the two ends of the system
are kept at different temperatures \cite{{Lebo},{Saito}}.
The temperature profile that forms
is flat everywhere apart from the neighborhood of the boundaries, and
the energy flux is clearly not proportional to the temperature gradient
inside the sample. The origin of this anomaly is the fact that
currents in integrable systems (energy currents in the above examples)
often emerge as integrals of motion thus causing  the
transport coefficients to be divergent or ill defined \cite{Zotos}.

The existence of flat density (temperature) profiles
in the presence of (energy) currents points to
the important role played by the homogeneous, current-carrying states
in integrable systems.
In previous papers \cite{{trising},{trxy}}, we
suggested that a method of constructing such states
at zero temperature is to add to the Hamiltonian
an appropriate current with a
Lagrange multiplier and then find
the ground state of the system. This method can be
applied with relative ease to models such as
the transverse Ising and isotropic $XY$ (so called $XX$) models, and one can
calculate the effect of energy current on experimentally
measurable correlations.

Since there are physical systems that are rather well
approximated as $XX$ models \cite{1drev}, and since the effect of
currents should be measurable in inelastic neutron scattering
experiments \cite{RZ}, one should carefully examine the validity of the
assumptions underlying the calculations of the correlations.
The Lagrange multiplier method is based on the assumption that
the homogeneous, current-carrying states are the same whether
they were induced by boundary conditions or by bulk driving
fields. This is not an obvious assumption and its validity
should be checked by first solving the boundary condition problem
and then comparing the results
with those obtained from the Lagrange-multiplier method.
Unfortunately, the execution of this program runs into difficulties
since the boundary condition problem is nontrivial and
we cannot provide an exact solution for any of the interesting cases.

One can make some progress, however, in a closely related problem.
Namely, current-carrying states
generated by inhomogeneous initial conditions can be studied
and one can ask whether those states were describable by the
Lagrange multiplier method. In order to illuminate
the new problem, let us imagine the original problem as finding
the steady state of a brick with two of its opposite sides kept at
different temperatures.
In the new problem, we have two semi-infinite bricks at different
temperatures and, at time zero, the bricks are joined at the ends.
If the bricks were integrable systems then one would expect that the
time-evolution would produce an interesting temperature
profile with a flat part around the joining point
and it would be meaningful to compare the profile as well as
various local expectation values to those obtained from the
Lagrange-multiplier method. This is what we shall do in this paper
for the $XX$ chain after making the following
simplifications. First, we note that
the transverse magnetization, $m$,
is a conserved quantity in the $XX$ model. Since this is a much simpler
quantity to treat, we shall study the transport of
$m$ instead of the transport of the energy. Second,
we restrict our study to the zero-temperature properties of the system.
Third, we shall use a simple inhomogeneous initial state
pasted together from two zero-temperature homogeneous pieces with
$m(x<0,0)=m_0$ and $m(x>0,0)=-m_0$.

As a result of the above simplifications,
the problem of time-evolution becomes solvable and the formation
of a flat magnetization profile can be observed.
We find that both the magnetization profile
and the fermionic two-point correlations agree
with the results obtained from the Lagrange multiplier method, thus
indicating that the method does work for obtaining the
homogeneous current carrying states.

We start (Sec.\ref{model}) by defining the system and constructing
the initial state in terms of fermionic operators.
Next, the time evolution of the
magnetization is calculated and the scaling form, $m(x,t)=\Phi(x/t)$,
of the long-time limit is discussed (Sec.\ref{magn}). Finally, we show that
the local properties of the scaling limit can be described
using the homogeneous, current carrying states obtained from the
Lagrange multiplier method (Sec.\ref{conserv}).

%%%%%%%%%%%%%%%%%%%%%%%%%%%%%%%%%%%%%%%%%%%%%%%%%%%%%%%%%%%%%%%%%%%%%

\section{The model and the construction of the initial state}
\label{model}

The system we investigate is the $XX$
model defined by the Hamiltonian
 \be
 H^{xx}=\sum_{n}h^{xx}_n 
 =-J\sum_{n} \left( s^x_n s^x_{n+1} + s^y_n s^y_{n+1}  \right)
 \label{trxyhami}
 \ee
where the spins, $s^\alpha_n$ ($\alpha =x,y,z$), are
$1/2$ times the Pauli matrices situated
at the sites of a one-dimensional
periodic lattice $[n=0,\pm 1, ...,\pm(N-1),N\ ; \
s^\alpha_{N+1}= s^\alpha_{-N+1}$].  The coupling, $J$,
is set to $J=1$ throughout this paper and
$N\to \infty$ is assumed in most of our calculations.

The $z$ component of the total magnetization, $M^z=\sum_n s^z_n$,
is conserved in this system and, using the continuity equation for $M^z$,
a local magnetization current can be defined \cite{trxy}:
 \be
 j^M_n=s^y_n s^x_{n+1} - s^x_n s^y_{n+1} \quad .
 \label{loccurrent}
 \ee
It can be easily shown that the global magnetization current,
$J^M=\sum_nj^M_n$, commutes with the Hamiltonian (\ref{trxyhami}) and thus
one can classify the eigenstates of $H^{\rm xx}$ by the
global current flowing through the system.
Below we shall study how the initial conditions can
generate states that are locally homogeneous
and carry a finite magnetization current.

The time-evolution that follows from a given initial condition
can be calculated for the $XX$ model since, using
a Jordan-Wigner transformation, this model can be transformed into a set
of free fermions \cite{LSM}
 \be
 H^{xx}=-\sum_k \cos k \ b^\dagger_k b_k
 \label{diaghxx}
 \ee
where $b_k$ is the annihilation operator of a fermion of
wave-number $k$. It follows that the time evolution of the $b_k$ operators
is known $ b_k(t) = b_k(0) \exp( i t \cos k) $
and a Fourier transform provides us with the time dependence of
a local Fermi operator, $c_n$, as a sum of
Bessel functions of the first kind \cite{Abram}
 \be
 c_n(t) = \sum_j c_j(0) i^{j-n} J_{j-n}(t) ~.
\label{c}
 \ee
The above expression is the starting point for our calculations
since we wish study the local transverse magnetization, $s_n^z$, that
can be expressed through the local Fermi operators as
 \be
 s^z_n = c^+_n c_n - {1\over2}  \ .
 \ee

Our aim is to find the time evolution of the expectation value of
$s_n^z$
 \be
 m(n,t)=\langle \varphi |s_n^z(t) |\varphi \rangle
 \ee
using an initial state $|\varphi\rangle$ that is expected to
produce a magnetization current.
The simplest such state has a symmetric
step in the initial magnetization
 \be
 \langle\varphi| s_n^z(0) |\varphi \rangle =
 \left\{\begin{array}{ll}\hskip 9pt m_0  ~~~ & \mbox{for $-N < n \le 0$} \cr
 -m_0  & \mbox{for \hskip 14pt $1\le n\le N$~.}
 \end{array} \right.
 \label{m_0}
 \ee
The above condition, however, does not specify the initial state uniquely.
In order to define $|\varphi \rangle$ precisely, we shall construct
the state of the $2N$ spins as a direct product of states of
two systems of $N$ spins. The {\it left} system contains the $n\le 0$ spins
while the {\it right} one is built from the rest ($n\ge 1$).
In order to get close to a zero-temperature, current-carrying state we
choose the lowest energy state under the above conditions (\ref{m_0}).
Namely, both half-chains are chosen to be in their
ground states with the magnetizations being $m_0$ and $-m_0$
in the {\it left} and {\it right}, respectively. Another
reason for choosing such a state comes
from trying to avoid possible energy currents in the system. In the
$N\to \infty$ limit, the above state
is a ground state of the model at fixed $m=\pm m_0$ and the two
homogeneous sides have the same energy. Thus one expects that
only magnetization current will flow in the course of the
time-evolution of the system.

A formal definition of $|\varphi \rangle$ with the above properties
can be given as follows. Let us define annihilation
operators on the {\it left} and on {\it right} by the
Fourier transforms of the
fermionic operators, $c_n$, on the appropriate sides of the system
 \be
 L_k = {1 \over \sqrt{N}} \sum_{j=-L+1}^0 e^{ikj} c_j \ ,\ \ 
 R_k = {1 \over \sqrt{N}} \sum_{j=1}^L e^{ikj} c_j \ .
 \ee
As can be shown, these operators are also fermionic operators
with the anticommutation relations:
$\{L_k, L_q^+\} = \{R_k, R_q^+\} = \delta_{kq}$, and
the initial state, $|\varphi\rangle$, can be written as
 \be
 |\varphi \rangle = \prod_{k=-k_-}^{k_-} R_k^+
           \prod_{k'=-k_+}^{k_+} L_{k'}^+ |\downarrow\,\,\rangle
 \label{leftright}
 \ee
where $|\downarrow\,\, \rangle$ is the state with all spins down and
$k_\pm =\pi(\frac{1}{2}\pm m_0)$. In the following $m_0>0$ will be assumed
without restricting the generality of the arguments.

%%%%%%%%%%%%%%%%%%%%%%%%%%%%%%%%%%%%%%%%%%%%%%%%%%%%%%%%%%%%%%%%%%%%%
\section{Numerical and analytical evaluation of the
transverse magnetization}
\label{magn}

Once the initial state $|\varphi\rangle$ is given,
the time evolution of  $s^z_n $ is expressed through initial
correlations
 \bea
 &&\langle \varphi| s^z_n(t) |\varphi \rangle =  \cr \cr
 &&\sum_{j,l} i^{j-l} J_{j-n}(t) J_{l-n}(t)
 \langle \varphi| c_l^+(0) c_j(0) |\varphi \rangle - {1\over2}~.
 \label{mzgeneral}
 \eea
It follows now from the special construction of $|\varphi\rangle$
that the terms with $j$ and $l$ on different side of the origin vanish.
Furthermore, the terms where
$l-j$ is odd cancel since the expectation value is real. The calculation
of $\langle \varphi| c_l^+(0) c_j(0) |\varphi \rangle$ is carried out
by using the explicit form
(eqs. \ref{leftright}) for $|\varphi\rangle$
and the result is as follows
\bea
 i^{j-s} \langle\varphi| c_s^+ &&c_j |\varphi \rangle = \cr \cr
 &&\left\{ \begin{array}{ll}
 \pm{\sin(\pi m_0(s-j)) \over \pi(s-j)}  ~~~ &
      \mbox{for $s-j$ even $\ne 0$} \cr
 {1 \over 2} \pm {m_0}  &
      \mbox{for $s-j = 0$} ~.\cr
 \end{array} \right.
 \label{expect}
\eea
The upper (lower) sign is valid for both $s$ and $j$
being in the {\it left} ({\it right}) part of the chain.
Since the above expression depends only on the even integer $s-j$,
the notation simplifies considerably by introducing $l=(s-j)/2$.
Treating the $l=0$ term separately and using an identity \cite{Abram}
for the Bessel functions, $\sum_n J_n^2 = 1$,
equations (\ref{mzgeneral}) and (\ref{expect}) yield
the following magnetization profile for $n\ge1$:
 \bea
 m(n,t) = -&& m_0 \sum_{j=1-n}^{n-1} J_j^2(t) \, - \cr \cr
 &&2 \sum_{l=1}^\infty {\sin 2 \pi m_0 l \over 2 \pi l}
 \sum_{j=1-n}^{n-1} J_j(t) J_{j+2l}(t) ~.
 \label{mzsum}
 \eea
The $n\leq 0$ values can also be calculated and they show
the symmetry of the initial state, $m(n,t)=-m(-n+1,t)$. The
above expression for $m(n,t)$ is the basis for both numerical
and analytical results presented below.

\subsection{Numerical results}

The infinite sum (\ref{mzsum}) defining $m(n,t)$ converges well
since the main contribution comes from terms with small $l$.
The numerical evaluation of $m(n,t)$
poses a problem only for $m_0\ll 1/2$ when
the scaling limit, $n\to \infty$, $t\to \infty$, with
$n/t\sim 1$, can not be readily reached due to the increasing
number of terms that must be taken into account when evaluating
the sums in (\ref{mzsum}).

\begin{figure}[htb]
	\centerline{
	\epsfxsize=0.45\textwidth
	\epsfbox{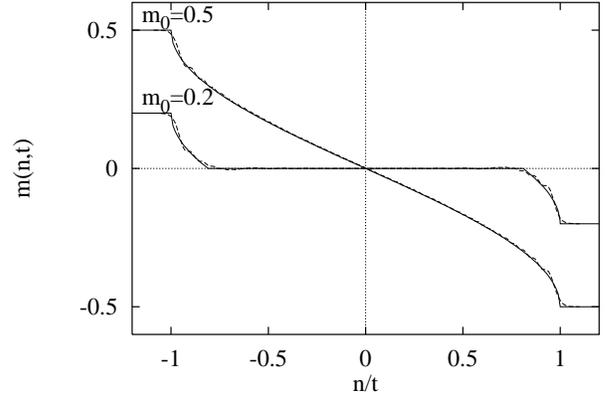}}
	\caption{Magnetization profile for initial 
	conditions $m_0=0.5$ and $m_0=0.2$. The large-time limits 
	approach the scaling curves (solid lines) 
	given by equation (\ref{scalingfunc}).}
	\label{fig:mscale}
	\end{figure}

The results from the numerical analysis (see Fig.\ref{fig:mscale})
show that the magnetization
profile evolves into a scaling form in the large-time limit
 \be
 m(n,t)\approx \Phi(n/t)
 \label{scaling}
 \ee
with the following scaling function (analytically obtained
in Sec.\ref{sec:anres}) giving an excellent fit to the data:
 \bea
 &&\Phi(v)=-\Phi(-v)= \cr \cr
 &&\left\{ \begin{array}{ll}
  0 & \mbox{for ~~~$0<v<\cos(\pi m_0) $} \cr
 -m_0+{1\over\pi}\arccos(v)~~~~ &\mbox{for ~~~$\cos(\pi m_0)<v<1 $} \cr
 -{1\over 2} & \mbox{for ~~~$1<v$} \cr
 \end{array} \right.
 \label{scalingfunc}
 \eea
As one can see, the magnetization profile does develop
a flat part that expands with a finite velocity
$c=\cos{\pi m_0}$. The velocity $c(m_0)$ decreases with increasing
$m_0$ and it diminishes for $m_0\to 1/2$.

\begin{figure}[htb]
	\centerline{
	\epsfxsize=0.45\textwidth
	\epsfbox{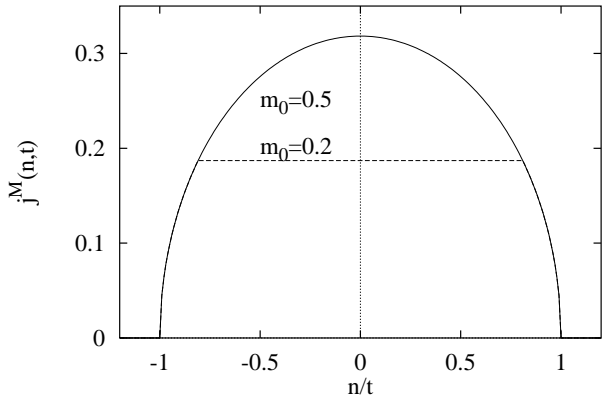}}
	\vspace{0.3truecm}
	\caption{Scaling functions for the magnetization-current for initial 
	conditions $m_0=0.5$ and $m_0=0.2$ as given by equation 
	(\ref{jscalingfunc}).}
	\label{fig:jscale}
	\end{figure}

The conservation of magnetization implies that the
magnetization current has a scaling form, $j^M(n,t)\approx \Psi(n/t)$,
as well, and the scaling function can be obtained from
(\ref{scalingfunc}) using the continuity equation together
with the boundary condition
$j^M(n\rightarrow\pm\infty,t)=0$. We find the scaling function
 \bea
 \Psi(v)=&&\Psi(-v)= \cr \cr
 &&\left\{ \begin{array}{ll}
 {1 \over \pi} \sin \pi m_0 ~~~~ &
     \mbox{for~~ $0<v<\cos(\pi m_0) $} \cr
 {1\over\pi} \sqrt{1-{v}^2} &
      \mbox{for~~ $\cos(\pi m_0)<v<1 $} \cr
 0 & \mbox{for~~ $1<v$} \quad ,\cr
 \end{array} \right.
 \label{jscalingfunc}
 \eea
shown in Fig.\ref{fig:jscale}.

\subsection{Analytical results}
\label{sec:anres}

Let us begin by deriving the analytical guess obtained from numerics
(\ref{scalingfunc}) by treating the case $m_0= 1/2$.
This limiting case is especially simple since $\sin{(2\pi m_0 l)} =0$ and
hence one is left with the following expression:
 \be
 m(n,t) = -\frac{1}{2}\sum_{j=1-n}^{n-1} J_j^2(t) = 
          -\frac{1}{2} + \sum_{j\ge n} J_j^2(t) ~.
 \label{mz1/2sum}
 \ee
We shall evaluate the scaling limit of
the above sum by calculating the integral of the
large $n$ limit of the discrete derivative
of $m(n,t)$ defined as\cite{discder}
 \be
 \label{fivn}
 \Phi'_n(v) =  t~[m(n+1,t)-m(n,t)]_{n/t=v}= 
  -\frac{n}{v}J_n^2\left(\frac{n}{v}\right) \ .
 \ee

The cases $v>1$ and $v<1$ must be treated separately since the asymptotic
properties of the Bessel functions change at $v=1$ \cite{Gradstein}. Namely,
one finds for $v>1$
 \be
 \Phi'_n(v) \sim {\frac{1}{2\pi\sqrt{v^2-1}}} 
 \exp\left[-\frac{2n}{v} (v\cosh^{-1}v -\sqrt{v^2-1})\right]
 \ee
while for $v<1 $
 \be
 \Phi'_n(v) \sim {2\over \pi \sqrt{1-v^2}}
 \cos^2 \left[\frac{n}{v}(\sqrt{1-v^2}-v\arccos{v}) - {\pi \over 4}\right]
 \label{blimit}
 \ee
As one can see,
the derivative approaches zero exponentially for $v>1$
(outside of the `light cone') and thus $\Phi(v)$ is constant
in this region. The constant is determined by the boundary condition
$\Phi(\infty)=-1/2$ and thus we have $\Phi(v)= -1/2$ for $v>1$.

The case $v<1$ is complicated by the fact that the limit
$n\to \infty$ does not exist for $\Phi'_n(v)$ due to the increasing
frequency of oscillations in (\ref{blimit}). It should be noted, however,
that we are interested in the integral of $\Phi'(v)$
 \be
 \label{osc}
 \Phi(v) =\lim_{n\rightarrow \infty} \Phi_n(v) = 
 \lim_{n\rightarrow \infty}
 \int_0^v\Phi'_n(y) dy \ ,
 \ee
and the $n\rightarrow \infty$ limit of the above
integral does exists. The value of the
integral can be obtained by replacing the fast oscillating $\cos^2$ term
[see (\ref{blimit})] by  its average value $1/2$. One then obtains
 \be
 \Phi(v) = -{1 \over \pi} \arcsin v ~,
 \ee
in agreement with (\ref{scalingfunc}) for $m_0=1/2$.

The evaluation of $\Phi(v)$ for $m_0\not= 1/2$ is more involved but,
essentially, it follows the above steps and yields
the general expression (\ref{scalingfunc}). An outline of the calculation
and some intermediate results are given in Appendix A.
%%%%%%%%%%%%%%%%%%%%%%%%%%%%%%%%%%%%%%%%%%%%%%%%%%%%%%%%%%%%%%%%%%%%%
\section{Comparison with the Lagrange multiplier method}
\label{conserv}

A consequence of the emergence of scaling, $m(n,t)\approx \Phi(n/t)$,
is that the magnetization becomes locally flat,
$m(n+1,t)-m(n,t) \sim t^{-1}\Phi'\to 0$, in the scaling limit.
Since these locally homogeneous segments of the system become
arbitrarily large for $t\to \infty$, one would like
to ask if they could be described as a homogeneous,
current-carrying state of an infinite chain as obtained
from the Lagrange multiplier method \cite{trxy}.

In order to formulate the question more precisely, let us note that,
for a given $n/t=v$ and in the neighborhood of $n$,
the scaling state of the system
may be characterized by the local values of the conserved
quantities and their fluxes, where the expectation value of a local operator
$a_n(t)$ is defined in the scaling limit as
 \be
 \la a \ra_v = \left. \lim_{n,t\rightarrow\infty}\la
 \varphi | a_n(t) | \varphi \ra \right|_{n/t=v} \ .
 \label{scalingaver}
 \ee
The simplest conserved quantities are the
magnetization, $m=\langle s^z \rangle_v=\Phi(v)$, and the energy density,
$\varepsilon=\langle h^{xx}\rangle_v$ with
the corresponding fluxes being the magnetization current,
$j^M=\langle j^M \rangle_v$, and the internal
energy flux, $j^I=\langle j^I \rangle_v$, where
 \be
 J^I =  \sum_n j^I_n = 
  \sum_n  s_n^z (s_{n-1}^y s_{n+1}^x - s_{n-1}^x s_{n+1}^y)\ .
 \ee
The energy density is homogeneous in the initial state
(apart from the local perturbation of the domain wall at the origin)
and it can be shown that this homogeneity remains intact in the long-time
limit. Thus $\varepsilon$ is
determined by the initial condition and, in the
scaling limit, can be expressed through $m$ and $j^M$
 \bea
 \varepsilon =  \varepsilon(m, j^M) = && -{1 \over \pi} \cos (\pi m_0) = \cr \cr
 && -{1 \over \pi}\cos\left[\pi |m| + \arcsin(\pi j^M)\right] \ .
 \label{emjm}
 \eea
Note that the homogeneity of the energy density implies
a vanishing energy flux, $j^I=0$, for all values of $v$ and so
equation (\ref{emjm}) can be viewed as an equation of state
connecting the relevant densities and fluxes.

It is clear that
$(m,j^M,\varepsilon,j^I)$ gives only
a partial characterization of the state since there are
infinitely many conservation laws in the $XX$ model. Nevertheless,
one may assume that $(m,j^M,\varepsilon,j^I)$ were sufficient to
describe {\em the states emerging in the scaling limit}. Then one can
ask if the properties of these states were the same as those
of the homogeneous, current-carrying state of an infinite system
where the values of $m,j^M,j^I$ were fixed by introducing
conjugate fields (Lagrange multipliers) and by finding
the ground state, $|\psi\rangle$, of the following Hamiltonian
\be
{\cal H}=H^{xx}-hM^z-\lambda_MJ^M-\lambda_IJ^I \ .
\label{newham}
\ee
The reason for this identification
is that the scaling emerges as an evolution from
an initial state of minimal energy at a given $m=m_0$.
We assume here that local states of the scaling limit remain
zero temperature states
in the sense that they have minimal energy density at the given values of
$m, j^M$, and $j^I$.

Since $m,j^M,j^I$ are set by adjusting the Lagrange multipliers,
a necessary condition for $|\psi\ra$ having the local
properties of the scaling states is the equality of its energy density as
a function of $m,j^M,j^I$ to the energy density $\varepsilon$ given
by (\ref{emjm}). Verification of this equality can be viewed as checking
whether the equation of states of the two systems were identical.

\begin{figure}[htb]
	\leftline{\hspace{-1.5cm}
	\epsfxsize=0.55\textwidth
	\epsfbox{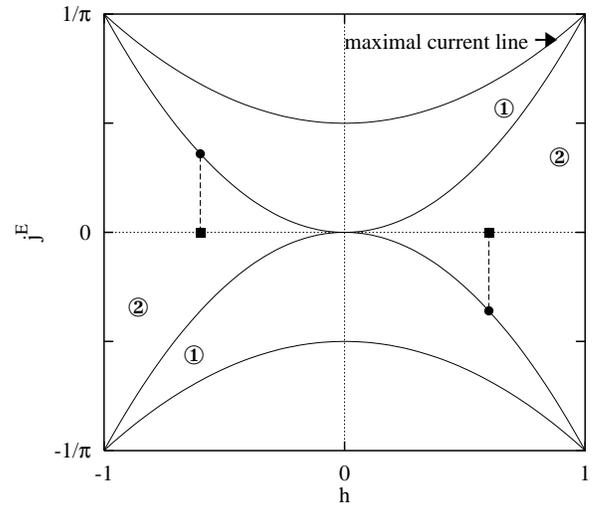}}
	\vspace{0.3truecm}
	\caption{Magnetic-field (h) -- energy-flux ($j^E=j^I-hj^M$) 
	phase diagram of an $XX$ model 
	where the energy flux is driven homogeneously in the bulk [5]. 
	The dashed lines in region \two correspond to states of constant 
	energy with the location of the dashed lines determined by the 
	initial conditions. The two filled squares represent
	the $m=\pm m_0$ regions while the filled circles describe identical 
	states 
	corresponding to the plato (m=0) in Fig.1. Moving along the dashed 
	lines 
	provides the local states $(m,j^M,j^I=0)$ in the transient region.}
	\label{fig:path}
	\end{figure}

In order to obtain $\langle\psi|h^{xx}_n|\psi\rangle$, we use
the results from our previous work \cite{trxy} where the
problem of energy current in the transverse $XX$ model was
investigated. In that work, we studied the ground-state
properties of the following Hamiltonian
 \be
 {\cal H}=H^{xx}-hM^z-\lambda_EJ^E \ .
 \label{newham2}
 \ee
with the total energy flux, $J^E$, being a sum of the fluxes of internal
and of magnetic energy
 \be
 J^E=J^I-hJ^M \ .
 \label{totalflux}
 \ee
We found that the $j^E,h$ phase diagram of this model has a region
(called region \two in Fig.1 of \cite{trxy}, see Fig.\ref{fig:path} of
the present paper) where the
ground-state expectation value of the flux of internal energy
is zero ($\langle \psi|J^I|\psi\rangle =0$) and
so the $j^I=0$ condition is
automatically satisfied. The magnetization flux
in region \two has been calculated in \cite{trxy}. For the case of
$h\lambda_E<0$ that is needed in the setup we use ($m_0>0$), it has the
following form
 \be
 j^M = -{j^E \over h} = {1\over\pi} \sqrt{1-\lambda_E^{-2}} \ ,
 \label{jm1}
 \ee
and the calculation of the densities of magnetization and of energy
is also a simple exercise with the following results
 \bea
 &&m = {1\over\pi} \left( \arcsin h -\arcsin \lambda_E^{-1} \right) -
 {1 \over 2}\hbox{sign}(h) \ ,
 \label{m1} \\ \cr
 &&\langle\psi|h^{xx}_n|\psi\rangle =
 -{1 \over \pi} \sqrt{1-h^2} \  . \label{eps1}
 \eea
Eliminating $h$ and $\lambda_E$ from (\ref{jm1},\ref{m1} and
\ref{eps1}) produces then (\ref{emjm}), i.e.
$\langle\psi|h^{xx}_n|\psi\rangle=\varepsilon (m,j^M)$.
This result indicates that
the local states derived as a scaling limit can indeed be
obtained with the help of the Lagrange method. 
Further confirmation
of the validity of this suggestion comes
from the studies of the scaling limit of the
fermionic number operator $\langle n_k \rangle_v$ in momentum space, which can
 be defined as follows:
\be
\langle n_k \rangle_v = \sum_m e^{-ikm}
\lim_{t\to \infty}\langle\varphi |c_{vt+m}^+(t)c_{vt}(t)|\varphi \rangle
\ee
This quantity can be calculated for the special case of
$m_0=1/2$ (see Appendix B) and one finds that
\bea
 \label{nk}
 &&\langle n_k \rangle_v= \cr\cr
 &&\left\{ \begin{array}{ll}
 {1 ~} &
     \mbox{if~ $-\arccos(v) +\pi/2 < k <\arccos(v) +\pi/2$},\cr
 0 & \mbox{else} \ .\cr
 \end{array} \right.
 \eea
Since $n_k$ is a projection operator with eigenvalues $0,1$ on the
eigenstates of $H^{xx}$ this result
shows that the system evolves locally (in the scaling sense described
by (\ref{scalingaver})) into an eigenstate of the $XX$ chain. The occupied
levels are the same as those obtained by the
Lagrange multiplier method for the given set of $(m,j^M,j^I=0)$.
This can be interpreted as the local scaling states being equivalent to the
 ground states found
in the Lagrange multiplier method.

In order to demonstrate the value of the equivalence suggested
by the above considerations,
let us show how simple is the derivation of the
magnetization profile if we assume that the Lagrange multiplier
method can be used to describe the scaling limit.

The calculation is based on the observation that the locally flat
magnetization $[m(n+1,t)-m(n,t)
\sim t^{-1}\Phi'\to 0]$ implies that one can take the continuum
limit of the equation describing the conservation of magnetization
\be
\partial_tm+\partial_x j^M=0 \ .
\label{continuumeq}
\ee
Next, using the fact that the energy density is constant in both space
and time one finds the states that can be used to
describe the scaling limit for a given initial magnetization.
It follows from eq.(\ref{eps1}) that these states are located along
$h=constant$ lines in the $j^E,h$
phase diagram as shown in Fig.\ref{fig:path}.
Along this path, the current $j^M$ depends only on the local magnetization
\be
j^M(x,t)=F(m)={1 \over \pi}\sin \left[ \pi(m_0-|m|)\right] \ ,
\label{j(m)}
\ee
where $F(m)$ is obtained from (\ref{jm1}) and (\ref{m1}) using the fact that
$h$ is given by the initial energy density through (\ref{eps1}).
It follows then that the continuity equation
(\ref{continuumeq}) has a scaling solution [including the solution $m=0$]
and that the magnetization profile, $m(x,t)=\Phi(x/t)$
can be obtained from the following equation
\be
\frac{dF}{dm}=\frac{x}{t} \ .
\label{magprof}
\ee
Taking into account the limiting behavior, $\Phi(z\to \pm \infty)=\pm m_0$,
the straightforward steps described above lead to the results
(\ref{scalingfunc}) obtained from a much
more complicated calculation treating the dynamics of the system.

%%%%%%%%%%%%%%%%%%%%%%%%%%%%%%%%%%%%%%%%%%%%%%%%%%%%%%%%%%%%%%%%%%%%%
\section{Final remarks}
\label{finalremarks}
We have shown that, in the zero-temperature $XX$ model, an 
initial state with a step-like magnetization inhomogeneity
evolves into a scaling state that is remarkable for its 
magnetization profile
(emergence of a flat part) and for the existence of locally 
homogeneous, current carrying states. We have also provided 
evidence that the local states of the 
scaling limit can be obtained by the Lagrange multiplier method. 
The possibility of applying the Lagrange method would simplify 
the calculation of transport significantly, thus it would be important 
to test the generality of our results. The simplest questions to  
ask are whether the results could be extended to finite temperatures and
whether the calculation could be generalized to the transport of energy 
as well. It appears that these questions can be 
answered and work is in progress along these lines. 
%%%%%%%%%%%%%%%%%%%%%%%%%%%%%%%%%%%%%%%%%%%%%%%%%%%%%%%%%%%%%%%%%%%%%

\section*{Acknowledgments}
This work has been partially supported by the 
by the Hungarian National Science Foundation 
(Grants OTKA T-019451 and T-017493).

%%%%%%%%%%%%%%%%%%%%%%%%%%%%%%%%%%%%%%%%%%%%%%%%%%%%%%%%%%%%%%%%%%%%%
\end{multicols}
\widetext

\appendix

\section{Magnetization profile}

We are considering the case $m_0 \neq 1/2$ and study the scaling limit of the
discrete derivative $\Phi'_n(v)$ defined in (\ref{fivn}):
\bea
\Phi'_n(v) &&= t\left[m(n+1,t) - m(n,t)\right]_{n/t=v} \cr \cr
&&= t\left[-{2m_0 \pi \over \pi}J_n^2(t) -2\sum_{r=1}^\infty{\sin(2m_0 \pi r)
\over 2\pi r} [J_n(t) J_{n+2r}(t) + J_{-n}(t)
J_{-n+2r}(t)]\right]_{n/t=v}
\eea
In order to take the scaling limit, we divide the sum into three parts where
the indexes of the Bessel functions are positive. Using the limiting
properties of the Bessel functions, one obtains the following form
($n\to\infty$ and $0<v<1$):
\bea
&&\Phi'_{n\to\infty}(v)=-{4m_0 \pi \over \pi^2\sqrt{1-v^2}}\cos^2 \left({n
\over v}\gamma(v)-{\pi \over 4}
\right) \cr \cr
&&-2\sqrt{2 \over \pi
\sqrt{1-v^2}}
\left(
\sum_{r=1}^\infty \left( A^+_r + A^-_r \right) -
 \left[1-(-1)^n \right] \sum_{r=\left[{n \over 2}\right]+1}^\infty A^-_r
\right) \cos\left({n \over v}\gamma(v)-{\pi \over 4} \right)
\eea
with
\bea
A^\pm_r=&&\lim_{n \to \infty}{\sin(2m_0 \pi r) \over 2\pi r}  \cos\left[{n \over v}\gamma\left(v\pm
{2rv \over n}\right) - {\pi \over 4}\right]
\sqrt{2\over \pi \sqrt{1-(v\pm 2rv/n)^2}} = \cr \cr
&&{\sin(2m_0 \pi r) \over 2\pi r} \cos\left[{n \over v}\gamma(v)-{\pi \over
4} -2r\arccos v \right] \sqrt{2\over \pi \sqrt{1-v^2}}
\label{oscsum}
\eea
where $\gamma(x)=\sqrt{1-x^2}-x\arccos x$. We can drop the second sum since it
vanishes with $n \to \infty$. 
Using trigonometrical identities, the calculation is reduced to evaluating the
following sums (for $0<x<2\pi$)
\be
\sum_{r=1}^\infty{\sin(rx) \over r} = {\pi-x \over 2} ~,~~~~\mbox{and}~~~~~
\sum_{r=1}^\infty{\cos(rx) \over r} = -{1 \over 2}\ln(2(1-\cos x)).
\ee
and one finds
\be
\label{loc}
\Phi'_{n\to\infty}(v)=
\left\{
\begin{tabular}{ll}
0	& ~~~~~for $v < \cos m_0 \pi$, \cr
$-{2 \over \pi\sqrt{1-v^2}}\cos^2({n \over v}\gamma(v)-\pi/4)$	& ~~~~~for $v >
\cos
m_0 \pi$.
\end{tabular}
\right.
\ee
Just as in Sec.~\ref{sec:anres} one can see that
$\Phi'\not=\lim_{n\to\infty}\Phi'_n$  since the differentiation and
the scaling limit cannot be exchanged. However, one can obtain $\Phi$ from
(\ref{loc}) by simply 
integrating it with the substitution $\cos^2 \rightarrow 1/2$ as
it is a rapidly oscillating function of $v$ in the scaling limit
\bea
\Phi(v) = &&\lim_{n\to\infty} \int^v \Phi'_n(y) dy = \cr \cr
&&\left\{
\begin{tabular}{ll}
0	& ~~~~~for $v < \cos m_0 \pi$, \\
$- \int_{\cos m_0 \pi}^v {dy \over \pi\sqrt{1-y^2}} = -m_0 +{1 \over
\pi}\arccos v$	& ~~~~~for $v > \cos m_0 \pi$.
\end{tabular} \right.
\eea

\section{Fermion occupation number in the scaling limit}

In a translationally invariant system the expectation value of the fermionic
number operator $\exval{n_k} = \exval{b^\dagger_kb_k}$ in momentum space
is expressed in terms of the real-space creation and annihilation operators
as the Fourier transform of the two-point correlation function
\be
\exval{n_k} = \sum_m e^{-ikm} \exval{c^\dagger_{j+m}c_j}
\ee
where the site number $j$ is arbitrary (because of translational invariance).
In order to investigate this expectation value in the scaling regime
we choose $j=vt$ and take the stationary limit $t \to \infty$ in the
expectation value {\em before} performing the summation over $m$.
In this way we probe the scaling state of the system in the region
of the shifted origin located in the point $vt$ where the state becomes
translationally invariant in the long-time limit.

At time $t$ the correlation function $\exval{c^\dagger_{j+m}c_j}$ for the
initial state with $m_0 =1/2$ is obtained from (\ref{c}) as
\bea
\exval{c^\dagger_{j+m}c_j}
 & = & (-i)^m \sum_{s=0}^\infty J_{s+m+j}(t) J_{s+j}(t) \nonumber \\
 & = & (-i)^m J_{m+j} J_{j} + \frac{(-i)^m t}{2 j}\left( J_{m+j} J_{j+1} -
J_{m+j+1} J_{j} \right).
\eea
Using the asymptotic expression of the Bessel functions \cite{Gradstein}
\be
J_{vt+m}(t) \sim \sqrt{\frac{2}{\pi t\sqrt{1-v^2}}}
\cos{\left(t\sqrt{1-v^2} - (vt+m)\alpha - \frac{\pi}{4}\right)}
\ee
with $\alpha = \arccos{v}$ and utilizing
the addition theorems for trigonometric functions we thus obtain
\bea
\lim_{t \to \infty}\exval{c^\dagger_{vt+m}(t)c_{vt}(t)} & = &
\frac{i^m}{\pi m} \sin{\alpha m} \nonumber \\
 & = & \frac{1}{2\pi} \int_{-\alpha + \pi/2}^{\alpha + \pi/2} dp \; e^{ipm}.
\eea
The Fourier transformation finally yields
\be
\exval{n_k}_v = \frac{1}{2\pi} \int_{-\alpha + \pi/2}^{\alpha + \pi/2} dp
\sum_m e^{i(p-k)m} = \int_{-\alpha + \pi/2}^{\alpha + \pi/2} dp \;
\delta(p-k)
\ee
and hence (\ref{nk}).

%%%%%%%%%%%%%%%%%%%%%%%%%%%%%%%%%%%%%%%%%%%%%%%%%%%%%%%%%%%%%%%%%
\begin{multicols}{2}
\narrowtext

\end{multicols}


\begin{references}
\bibitem{discder} It should be
noted that although $\Phi'(v)\not= \lim_{n\to \infty}\Phi'_n(v)$
the integral of $\Phi'_n(v)$ does converge to $\Phi(v)$.
\end{references}
\end{document}